\documentclass[12pt]{article}
\usepackage[mathscr]{eucal}
\usepackage{amssymb,latexsym}
\usepackage{verbatim}
\usepackage{amsmath}
\usepackage{amsthm}
\usepackage{graphicx}
\usepackage{cite}

\usepackage{url}
\usepackage{color}

\usepackage[normalem]{ulem}

\DeclareFontFamily{OT1}{rsfs}{}
\DeclareFontShape{OT1}{rsfs}{m}{n}{ <-7> rsfs5 <7-10> rsfs7 <10-> rsfs10}{}
\DeclareMathAlphabet{\mycal}{OT1}{rsfs}{m}{n}

\newcommand{\Nmanif}{{ M}}
\newcommand{\Mmanif}{{\mycal{S}}}
\newcommand{\red}[1]{\textcolor{red}{#1}}

\newtheorem{thm}{\sc Theorem}
\newtheorem{lem}[thm]{\sc Lemma}
\newtheorem{example}[thm]{\sc Example}

\newtheorem{prop}[thm]{\sc Proposition}
\newtheorem{rem}[thm]{\sc Remark}
\newtheorem{cor}[thm]{\sc Corollary}

{\catcode `\@=11 \global\let\AddToReset=\@addtoreset}

\newcounter{mnotecount}[section]

\renewcommand{\themnotecount}{\thesection.\arabic{mnotecount}}

\newcommand{\mnote}[1]
{\protect{\stepcounter{mnotecount}}$^{\mbox{\footnotesize
$
\bullet$\themnotecount}}$ \marginpar{
\raggedright\tiny\em
$\!\!\!\!\!\!\,\bullet$\themnotecount: #1} }

\renewcommand\S{\Sigma}

\renewcommand\d{\partial}

\renewcommand\th{\theta}

\newcommand\beq{\begin{eqnarray}}
\newcommand\eeq{\end{eqnarray}}
\newcommand\ben{\begin{enumerate}}
\newcommand\een{\end{enumerate}}
\newcommand\bit{\begin{itemize}}
\newcommand\eit{\end{itemize}}

\title{Positive mass theorems for asymptotically hyperbolic Riemannian manifolds with boundary\thanks{Preprint UWThPh 2021-9}}

\author{Piotr T.\ Chru\'{s}ciel\thanks{ email:
piotr.chrusciel@univie.ac.at; \protect\url{http://homepage.univie.ac.at/piotr.chrusciel}}
\\
Department of Physics\\ University of Vienna\\
$~$
\\
 Gregory J. Galloway\thanks{email: galloway@math.miami.edu}
 \\Department of Mathematics\\ University of Miami
}

\renewcommand{\red}[1]{#1}
\begin{document}
\date{}
\maketitle

\begin{abstract}
We prove positive mass theorems for asymptotically hyperbolic and asymptotically locally hyperbolic Riemannian manifolds with black-hole-type boundaries.
\end{abstract}


An interesting global invariant of asymptotically hyperbolic general relativistic initial data sets is the energy-momentum vector $\mathbf{m} \equiv (m_\mu)$
 \cite{ChHerzlich,Wang,ChNagyATMP}
 (compare~\cite{AbbottDeser,ChruscielSimon}).
It is known that $\mathbf{m}$ is timelike future pointing or vanishing~\cite{ChDelayHPET,HuangJangMartin} for conformally  compactifiable manifolds with
a connected conformal boundary at infinity with spherical topology and satisfying a natural lower bound on the scalar curvature.
The object of this note is to point out how to generalise this result to manifolds with a compact boundary
 and with several ends in dimensions $3\le n \le 7$, using the results of \cite{Eichmair:2020wbj,ChDelayHPET}, without assuming that the manifold $\Nmanif$ carries a compatible spin structure:

\smallskip

\begin{thm}
 \label{T26V21.1}
Let $(\Nmanif,h)$ be a conformally compact
  $n$-dimensional, $3\le n \le 7$,  asymptotically locally hyperbolic   manifold  with boundary.
Assume that the scalar curvature  of $\Nmanif$ satisfies $R(h) \ge  -n(n-1)$, and that the boundary has mean curvature
 $H \le n-1$
 with respect to the normal pointing into $M$.  Then,   the energy-momentum vector
 ${\bf m}$ of every spherical component of the conformal boundary at infinity of $(\Nmanif,h)$ is future causal.
\end{thm}

\begin{rem}
{\rm
Neither the boundary $\d \Nmanif $, nor the conformal boundary at infinity of $\Nmanif$, need to be connected.
}
\qed
\end{rem}

 Our sign convention for the mean curvature is such that the round sphere in Euclidean space has {\it positive} mean curvature with respect to the {\it outward} pointing normal.

%
%
%

In order to avoid ambiguities, some definitions are in order.
We say that a Riemannian manifold $(\Nmanif,h)$ is \emph{conformally compact} if there exists a compact manifold with boundary $\widehat  \Nmanif$ such that the following holds:
First, we allow $\Nmanif$ to have a boundary, which is then necessarily compact. Next,
$\Nmanif$ is the interior of $\widehat  \Nmanif$, whose boundary is the union of the boundary of $\Nmanif$ and of a number of new boundary components, at least one, which form the conformal boundary at infinity. Next, there exists on $\widehat  \Nmanif$ a smooth function $\Omega\ge 0 $  which is positive on $\Nmanif$, and which vanishes precisely on the new boundary components of $\widehat  \Nmanif$, with $d\Omega$ nowhere vanishing there. Finally,   the tensor field $\Omega^2 h$ extends to a smooth metric on $\widehat  \Nmanif$.

  An \emph{asymptotically locally hyperbolic} (ALH) metric is a metric with all sectional curvatures approaching minus one as the conformal boundary  at infinity is approached.
An \emph{asymptotically hyperbolic} (AH) metric is an ALH metric with spherical conformal boundary at infinity.

\begin{rem}
{\rm
 For simplicity our definitions require  smoothness up-to-boundary of the conformally rescaled metric, though much weaker conditions suffice for the result.
}
\qed
\end{rem}

\begin{example}
 \label{Ex26VI21.1}
{\rm
An example to keep in mind is provided by the space part of the Birmingham-Kottler metrics,
%
\begin{equation}\label{26XI21.1}
  g = V^{-2} dr^2 + r^2 h_{-k}
  \,,
  \qquad
  V^2 = r^2 + k - \frac{2m}{r^{n-2}}
   \,.
\end{equation}
where $h_{-k}$ is an Einstein metric on an $(n-1)$ dimensional manifold with scalar curvature equal to $ k (n-1)(n-2)$. The field of normals to the level sets of $r$ equals $N= V \partial_r$, hence
\begin{equation}\label{26XI21.2}
  H = \frac{1}{\sqrt{\det  g}} \partial_i (\sqrt{\det  g} N ^i)=
  \frac{V}{r^{n-1}} \partial_r (r^{n-1})
  =  (n-1)\sqrt{1 + k r^{-2}  -  2m r^{-n}}
  \,.
\end{equation}
When $k=1$ we find $H > (n-1)$  (with respect to the outward normal) for all sufficiently large $r$  regardless of the sign of $m$.

When $m<0$ the range of $H$ covers the interval $ (n-1,\infty)$ when $r$ runs over $(0,\infty)$, which shows that the condition on $H$ in Theorem~\ref{T26V21.1} is optimal.
\qed
}
\end{example}

In \cite[Theorem~1.1]{CGNP} we proved  non-negativity
 of mass for  ALH manifolds whose closure has
  topology $[0,1]\times (S^{n-1}/\Gamma)$, where $\Gamma$ is a finite subgroup of $SO(n-1)$. In particular  it was assumed that both conformal infinity and the boundary have the same topology $S^{n-1}/\Gamma$. Such manifolds will be referred to as \emph{ALH manifolds with product topology $[0,1]\times (S^{n-1}/\Gamma)$}. An example is provided by the Birmingham-Kottler metrics \eqref{26XI21.1}. Theorem~\ref{T26V21.1} allows us to prove the following improvement of this result:

\begin{cor}
 \label{C30VI21.1}
 Under the remaining conditions of Theorem~\ref{T26V21.1}, suppose instead that
$\Nmanif$ is a connected sum of a finite number (possibly zero) of closed manifolds, and a finite number (possibly zero) of conformally compact ALH manifolds, and of an ALH manifold with product topology $[0,1]\times (S^{n-1}/\Gamma)$, where $\Gamma$ is a finite subgroup of $SO(n-1)$. Then the mass of this
last component of the boundary at infinity is non-negative.
\end{cor}

\noindent{\sc Proof of Corollary~\ref{C30VI21.1}.}  For simplicity, consider the special case,
$$
{\widehat  \Nmanif} \cong \left([0,1]\times (S^{n-1}/\Gamma )\right) \# Q  \,.
$$
where $Q$ is a closed manifold.  We now make use of the following fact which is established by simple cut and paste arguments.

\smallskip
\noindent
\textsc{Fact.}
Let $X$ and $Y$ be $n$-manifolds.
Suppose $X'$ is a finite cover of $X$ with $r > 1$ sheets.  Then
$$
X' \# \underbrace{Y \# \dotsb \#Y}_{r\: {\rm times}} \quad \text{ is a cover of } \quad X \#Y \,.
$$

\smallskip
It follows that $(\Nmanif,h)$ admits a finite Riemanian cover of degree $r = |\Gamma|$, $({\Nmanif}',h')$, such that
$$
\widehat{{\Nmanif}'} \cong \left( [0,1] \times S^{n-1} \right) \# \underbrace{Q \# \dotsb \#Q}_{r\: {\rm times}}  \,.
$$

We may now apply Theorem~\ref{T26V21.1} to $({\Nmanif}',h')$.   The general case is handled in a similar manner.\qed

\smallskip
\begin{rem}
{\rm
In  \cite[Theorem~1.1]{CGNP}, we also proved positivity of mass for  ALH manifolds $\Nmanif$ with (nonempty) boundary so that $\widehat  \Nmanif$ has product topology $[0,1]\times T^{n-1}$.  By using Theorem 1.2 in \cite{Eichmair:2020wbj}, this can be improved in several respects. First, the condition $H< n-1$ can be replaced by $H\le n-1$. Next,  the assumption that  $\d \Nmanif \cong T^{n-1}$ can be replaced by the assumption $\d \Nmanif$ satisfies what is called  the {\it cohomology condition} in \cite{Eichmair:2020wbj}. This insures, in particular, that $\d \Nmanif$ cannot carry a metric of positive scalar curvature; see \cite[Theorem 2.28]{Lee:book}, compare~\cite[Theorem~5.2]{SchoenYau2017}.
Finally, the assumption, that
$\widehat {\Nmanif}$ has product topology, can be weakened to the assumption that  $\widehat {\Nmanif}$ satisfies the {\it homotopy condition} with respect to $\d \Nmanif$, as defined in \cite{Eichmair:2020wbj}.  This condition is satisfied if, in particular, there exists a retract of $\widehat {\Nmanif}$ onto $\d \Nmanif$.  A special case of \cite[Theorem 1.2]{Eichmair:2020wbj} is used in the proof of Theorem  \ref{T26V21.1}; see Proposition~\ref{global.foliation} below.

As already pointed-out in \cite[Remark~6.1]{CGNP},
the Horowitz-Myers solutions admit a CMC foliation with tori whose mean curvature approaches $n-1$ from above. This shows that the condition $H \le n-1$ is also sharp in the toroidal case.
}
\qed
\end{rem}

\smallskip

{\noindent\sc Proof of Theorem~\ref{T26V21.1}.}
 We note first that three dimensional manifolds are spin, and this case is already covered in \cite{ChHerzlich}.
Hence it remains to assume that $n\ge 4$.

Next, the proof is reduced to the case of a single conformal boundary at infinity with spherical topology by checking, via a calculation similar to \eqref{26XI21.2}, that the mean curvature, with respect to the  normal pointing towards the nearby conformal boundary at infinity, of the level sets of the conformal factor
$\Omega$, approaches $n-1$ as the conformal boundary at infinity is approached. We can therefore cut away an asymptotic end    of  $\Nmanif$
by introducing a new boundary component $\{\Omega=\epsilon\}$, with $\epsilon$ sufficient small so that this new boundary component satisfies, say, $H> 0$ with
  respect to the outward normal (thus $H<0<n-1$ with respect to the inward normal). This boundary component will be made part of the boundary of the new, truncated, manifold, still denoted by $\Nmanif$.

Summarising, we only need to prove the result for dim$\,\Nmanif \ge 4$ and with a conformally compactifiable ALH manifold $(\Nmanif,h)$ with a connected spherical conformal boundary at infinity.

Finally, we will need the following lemma. Its hypotheses are tailored towards obtaining a contradiction: we will show that no well-behaved AH manifolds satisfying the hypotheses of the lemma exist.

\begin{lem}
\label{chrudelay}
Let
 $(\Nmanif,h)$ be an
 $n$-dimensional, $n \ge 4$, ALH manifold with boundary.
Assume that the scalar curvature  of $\Nmanif$ satisfies $R(h) \ge  -n(n-1)$, and that the boundary has mean curvature $H \le n-1$ with respect to the normal pointing into $\Nmanif$.   Then, if the energy-momentum vector
 ${\bf m}$ of an AH end of $(\Nmanif,h)$ is spacelike or past causal, there exists a general relativistic initial data set $(\Mmanif,g,K)$ with boundary, dim~$ \Mmanif =$ dim $\Nmanif$, with the following properties.
\ben
\item $(\Mmanif,g,K)$ satisfies the Dominant Energy Condition (DEC).
\item  The boundary $\d \Mmanif =S_0$  satisfies $\th^+ \ge 0$ with respect to the normal  pointing  out of $\Nmanif$.
\item There exists an open  set $U$ containing $S_0$ and bounded away from the asymptotically hyperbolic end under consideration, such that outside $U$, $g = g_E$  (Euclidean metric) and $K = 0$.
\een
\end{lem}

\noindent{\sc Proof of Lemma~\ref{chrudelay}:}
The  proof of~\cite[Theorem~1.5]{ChDelayHPET}
provides an asymptotically hyperbolic manifold $(\Mmanif ,g_1)$ with $R(g_1)\ge -n(n-1)$
and with timelike past-directed energy-momentum vector.    As the manifold
$\Mmanif $ arises as a certain doubling  of $\Nmanif$ (via a localised ``Maskit gluing'' procedure), it contains two copies of the boundary  $\d \Nmanif $, with the metric near each copy of $\d \Nmanif $ coinciding with the original one, in particular the inequality $H \le n-1$ still holds. Using a deformation of the metric  near the conformal boundary at infinity  as in \cite[Corollary~1.4]{CGNP} one obtains a new asymptotically hyperbolic metric $g_2$ on $\Mmanif $ with constant negative mass aspect function and with $R(g_2)\ge -n(n-1)$.

To continue we  use the deformation arguments from \cite{AnderssonGallowayCai}. 
Indeed, by ~\cite[Theorem~3.2]{AnderssonGallowayCai} there exists a constant $a\in (0,1)$,  which we will denote by $a^2$ here, and
a metric $\widehat  g$ on $\Mmanif $ which coincides with the original metric near both copies of the original boundary $\d \Nmanif $, and which coincides with the metric
$$
 \frac{dr^2}{1+\frac{r^2}{a^2}} +r^2 h_0
 \,,
$$
for large $r$,
where $h_0$ is the unit round metric on $S^{n-1}$. Moreover the metric $\widehat  g$ satisfies $R(\widehat  g) \ge -n(n-1 )/a^2$ everywhere.

Let $\widehat  g_a = a^{-2} \widehat  g$, then $R(\widehat  g_a) \ge -n(n-1 )$. Let $N$, respectively  $\widehat  N_a$, be the $g$-unit, respectively the $\widehat  g_a$-unit normal to $\d \Nmanif  $, and let $\widehat  H_a$ denote the mean curvature of $\d \Nmanif$ in the metric $\widehat  g_a$. Then
$\widehat  N_a =  a N$ and
\begin{equation}\label{25VI21.1}
  \widehat  H_a = \frac{1}{\sqrt{\det \widehat  g_a}} \partial_i (\sqrt{\det \widehat  g_a} \widehat  N_a ^i)= aH \le a (n-1) \le n-1
  \,.
\end{equation}
 As in the proof of~\cite[Theorem~1.4]{ChDelayHPET}, view the exact hyperbolic end of $(\Mmanif,\widehat  g_a)$ as embedded in Minkowski space.
 By flattening out this hyperboloidal end (outside a sufficiently large sphere) within Minkowski space, one obtains a  general relativistic initial data set $(\Mmanif',\widehat g_a',\widehat  K_a')$, with  $\widehat K_a' = \widehat  g_a$ on the unchanged part of $\Mmanif$, which satisfies the requirements of the lemma.
\qed

\medskip

\begin{figure}
  \centering
  \includegraphics[width=.4\textwidth]{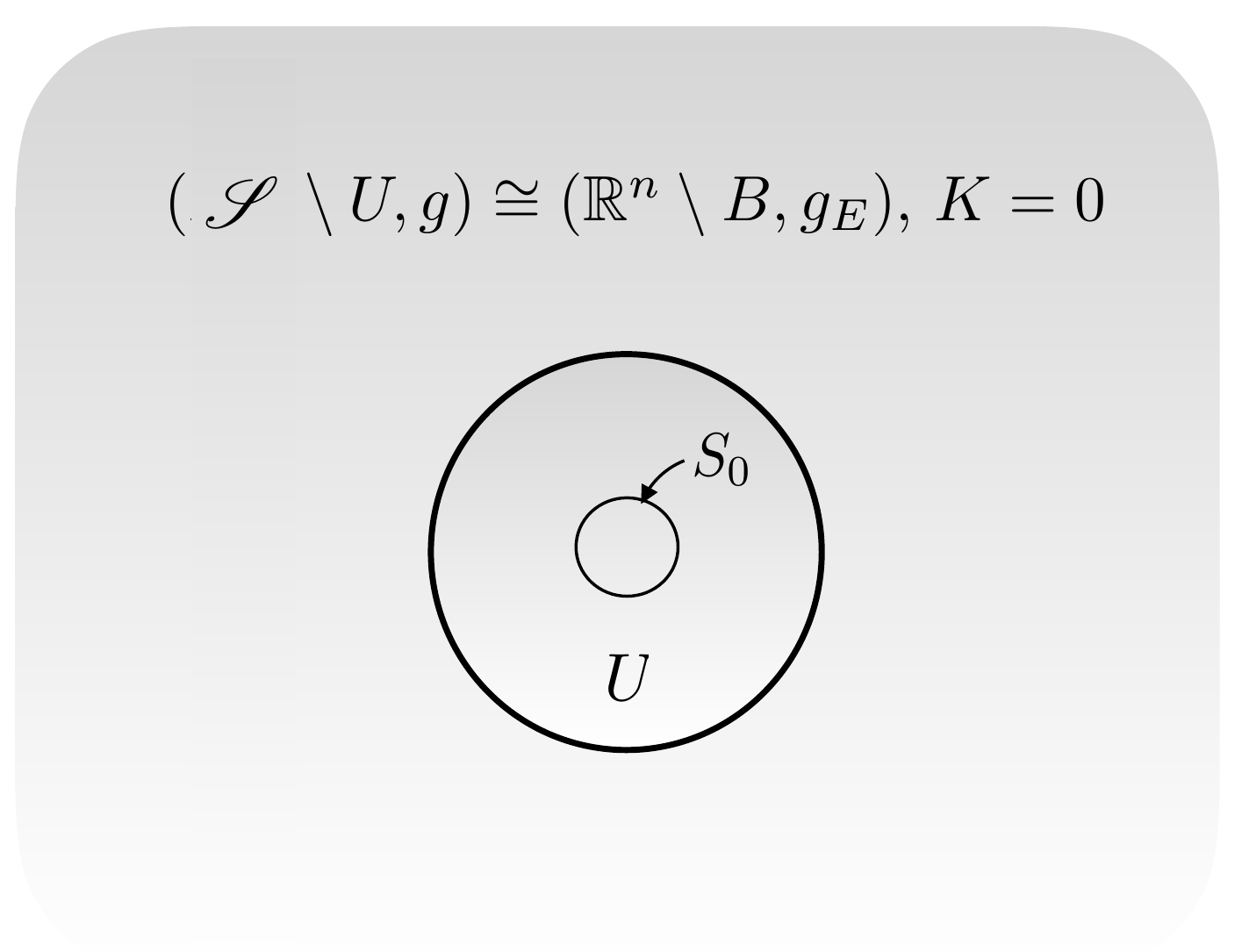}\qquad
  \includegraphics[width=.4\textwidth]{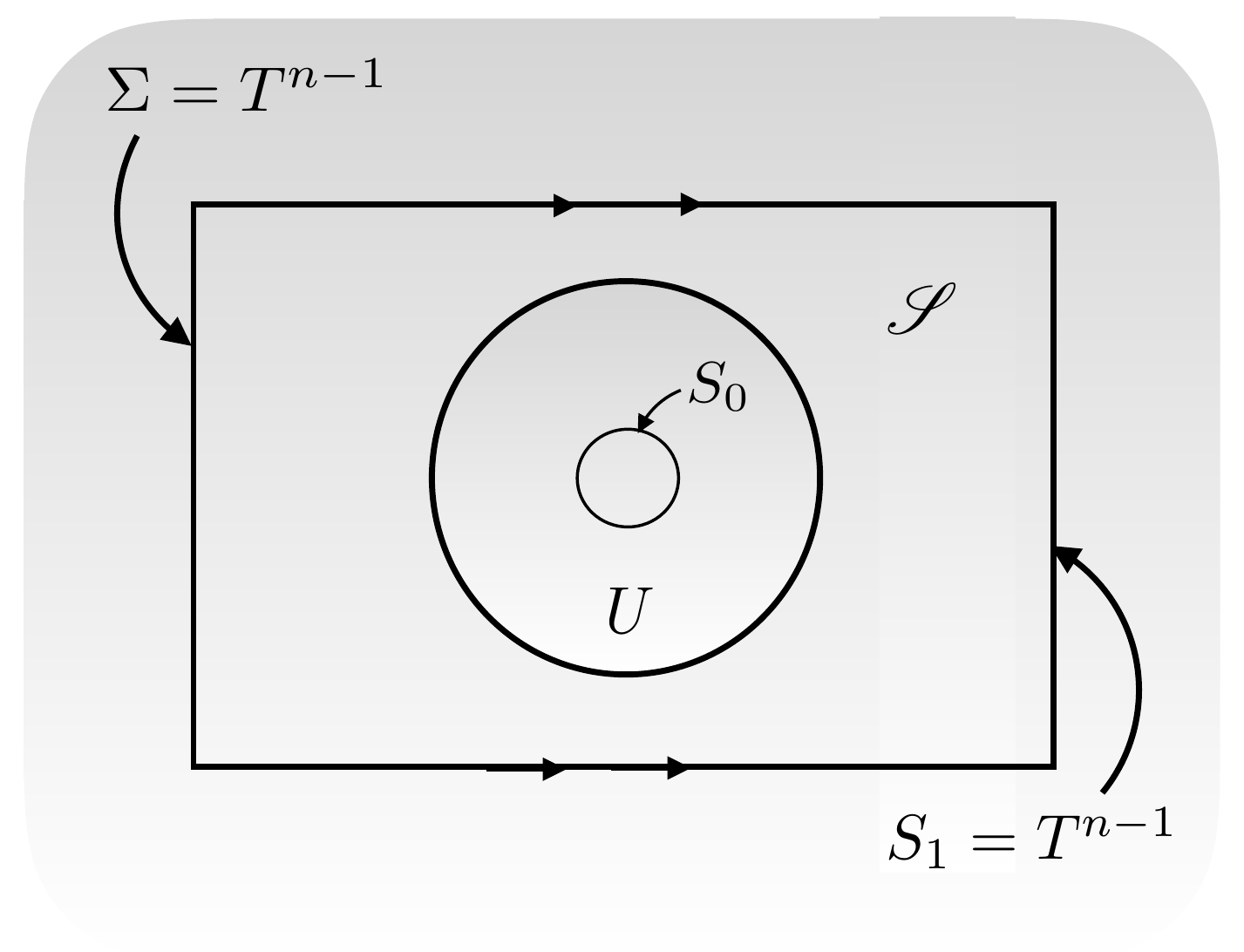}
  \caption{Left figure: The initial manifold. Right figure: The boundary of $M$ consists of two tori, $\{\pm R\}\times T^ {n-1}$ and the original boundary $S_0$.}\label{F15VI21.1}
\end{figure}

 We return to the proof of Theorem~\ref{T26V21.1}.
Suppose to the contrary that ${\bf m}$ is spacelike or past pointing causal.  Then by Lemma~\ref{chrudelay}, there exists an initial data set $(\Mmanif,g,K)$ with the properties listed in the theorem. Moreover, since for the proof  we are assuming that $(\Nmanif,h) $ is conformally compactifiable, by enclosing $\overline U$ inside a large  box $\partial([-R,R]^n)$, and then identifying  $n-1$ parallel sides of the box as in Figure~\ref{F15VI21.1},
we obtain an initial data set \emph{with boundary}, which we still refer to as $(\Mmanif,g,K)$, where
$(\Mmanif,g,K)$ is a compact manifold with boundary of the form
\begin{equation}\label{16VI21.11}
\Mmanif \cong \big( [0,1] \times T^{n-1} \big) \# Q
\,,
\end{equation}
where $Q$ is a compact manifold with boundary,
with the following additional properties.

\ben
\item $(\Mmanif,g,K)$ obeys the DEC.
\item $\d \Mmanif = \S \cup S$,  $S = S_0 \cup S_1$, where $S_0$ is weakly outer trapped, and
$\S$ and $S_1$ are flat  $(n-1)$-tori, each of which has   both null second fundamental forms equal to zero.   In particular, both are MOTS with respect to either normal, $\th^{\pm} = 0$.
\een

Such an initial data set is incompatible with the following special case of Theorem~1.2 in \cite{Eichmair:2020wbj}:

\begin{prop} \label{global.foliation}
Let $\red{{(\Mmanif,g,K)}}$ be an $n$-dimensional,  $3 \leq n \leq 7$, compact-with-boundary initial data set.  Suppose that $\red{{(\Mmanif,g,K)}}$ satisfies the DEC and that  the boundary $\d \red{ \Mmanif }$  can be expressed as a disjoint union of hypersurfaces,
$\d \red{ \Mmanif } =$ $\S\cup S$, such that the following conditions are satisfied:

\begin{enumerate}
\item $\theta^+\le0$ along $\S$ with respect to the normal pointing into $\red{ \Mmanif }$, and
$\theta^+\ge 0$ along $S$ with respect to the normal pointing out of $\red{ \Mmanif }$,
\item $\S \cong T^{n-1}$ and there exists a retract of $\red{ \Mmanif }$ onto $\S$.
\end{enumerate}

Then $\red{ \Mmanif }$ is diffeomorphic to  $[0,1]\times T^{n-1}$.  In particular, $S$ is connected.
\qed
\end{prop}

The constructed initial data set $(\Mmanif,g,K)$ satisfies all the assumptions of the proposition.  In particular from the `almost product' structure \eqref{16VI21.11} of $M$ it is easy to see that a retract of $M$ onto $\S$ exists.  Indeed, by suitably ``collapsing''  $Q,$ we obtain a continuous map from the right hand side of \eqref{16VI21.11} to  $[0,1] \times  T^{n-1}$, and use the product structure to project this to $\{0\} \times T^{n-1}$. We then reach a contradiction since $S$ is not connected.
\qed

\smallskip
\begin{rem}
{\rm
  Even under more general assumptions, Theorem 1.2 concludes much more.  In particular, in suitable coordinates, each slice $\S_t \cong \{t\} \times T^{n-1}$ is flat with vanishing null second fundamental form, $\chi^+ = 0$.
}
\qed
\end{rem}

\smallskip
\begin{rem}
{\rm
We have chosen to carry out the argument using an embedding of the hyperbolic end into Minkowski spacetime, with an eye out on possible  dimensional generalisations, taking advantage of the analysis in~\cite{Lohkamp2,SchoenYau2017}. In the dimensions considered here one can, instead, use a compactification argument similar to what is done in~\cite{AnderssonGallowayCai}, together with our remaining arguments, to achieve the same conclusions.
}
\qed
\end{rem}

\bigskip

{\noindent \sc Acknowledgements} The research of PTC was supported in part by a grant FWF P 29517-N27,
and by a grant of the Polish National Center of Science (NCN) 2016/21/B/ST1/00940. The research of GG was supported in part by an NSF grant DMS-1710808.

\providecommand{\bysame}{\leavevmode\hbox to3em{\hrulefill}\thinspace}
\providecommand{\MR}{\relax\ifhmode\unskip\space\fi MR }
\providecommand{\MRhref}[2]{%
  \href{http://www.ams.org/mathscinet-getitem?mr=#1}{#2}
}
\providecommand{\href}[2]{#2}


\begin{thebibliography}{10}

\bibitem{AbbottDeser}
L.F. Abbott and S.~Deser, \emph{Stability of gravity with a cosmological
  constant}, Nucl.\ Phys. \textbf{B195} (1982), 76--96.

\bibitem{AnderssonGallowayCai}
L.~Andersson, M.~Cai, and G.J. Galloway, \emph{Rigidity and positivity of mass
  for asymptotically hyperbolic manifolds}, Ann.\ H.~Poincar\'e \textbf{9}
  (2008), 1--33, arXiv:math.dg/0703259. \MR{MR2389888 (2009e:53054)}

\bibitem{ChDelayHPET}
P.T. Chru\'{s}ciel and E.~Delay, \emph{{The hyperbolic positive energy
  theorem}},  (2019), arXiv:1901.05263 [math.DG].

\bibitem{CGNP}
P.T. Chru\'{s}ciel, G.J. Galloway, L.~Nguyen, and T.-T. Paetz, \emph{{On the
  mass aspect function and positive energy theorems for asymptotically
  hyperbolic manifolds}}, Class. Quantum Grav. \textbf{35} (2018), 115015,
  arXiv:1801.03442 [gr-qc].

\bibitem{ChHerzlich}
P.T. Chru\'{s}ciel and M.~Herzlich, \emph{The mass of asymptotically hyperbolic
  {R}iemannian manifolds}, Pacific Jour.\ Math. \textbf{212} (2003), 231--264,
  arXiv:math/0110035 [math.DG]. \MR{MR2038048 (2005d:53052)}

\bibitem{ChNagyATMP}
P.T. Chru\'{s}ciel and G.~Nagy, \emph{The mass of spacelike hypersurfaces in
  asymptotically {anti -- de Sitter} spacetimes}, Adv.\ Theor.\ Math.\ Phys.
  \textbf{5} (2001), 697--754, arXiv:gr-qc/0110014.

\bibitem{ChruscielSimon}
P.T. Chru\'{s}ciel and W.~Simon, \emph{Towards the classification of static
  vacuum spacetimes with negative cosmological constant}, Jour.\ Math.\ Phys.
  \textbf{42} (2001), 1779--1817, arXiv:gr-qc/0004032.

\bibitem{Eichmair:2020wbj}
M.~Eichmair, G.J. Galloway, and A.~Mendes, \emph{{Initial data rigidity
  results}}, Commun.\ Math.\ Phys. (2020), Online First; arXiv:2009.09527
  [gr-qc].

\bibitem{HuangJangMartin}
L.-H. Huang, H.C. Jang, and D.~Martin, \emph{{Mass rigidity for hyperbolic
  manifolds}}, Commun.\ Math.\ Phys. (2019), 1--21, arXiv:1904.12010 [math.DG].

\bibitem{Lee:book}
D.A. Lee, \emph{Geometric relativity}, Graduate Studies in Mathematics, vol.
  201, American Mathematical Society, Providence, RI, 2019. \MR{3970261}


\bibitem{Lohkamp2}
J.~{Lohkamp}, \emph{{The Higher Dimensional Positive Mass Theorem II}},
  (2016), arXiv:1612.07505 [math.DG].


\bibitem{SchoenYau2017}
R.~Schoen and S.-T. Yau, \emph{{Positive Scalar Curvature and Minimal
  Hypersurface Singularities}},  (2017), arXiv:1704.05490 [math.DG].

\bibitem{Wang}
X.~Wang, \emph{Mass for asymptotically hyperbolic manifolds}, Jour.\ Diff.\
  Geom. \textbf{57} (2001), 273--299. \MR{MR1879228 (2003c:53044)}

\end{thebibliography}
\end{document}